\documentclass[fleqn,usenatbib]{mnras}

\usepackage{newtxtext,newtxmath}

\usepackage[T1]{fontenc}
\usepackage{ae,aecompl}

\DeclareRobustCommand{\VAN}[3]{#2}
\let\VANthebibliography\thebibliography
\def\thebibliography{\DeclareRobustCommand{\VAN}[3]{##3}\VANthebibliography}

\usepackage{graphicx}	

\newcommand{\km}{\rm\thinspace km}

\newcommand{\erg}{\rm\thinspace erg}
\newcommand{\s}{\rm\thinspace s}
\newcommand{\Mpc}{\rm\thinspace Mpc}

\newcommand{\kmps}{\hbox{$\km\s^{-1}$}}
\newcommand{\ergps}{\hbox{$\erg\s^{-1}\,$}}

\newcommand{\Msun}{\hbox{$\rm\thinspace M_{\odot}$}}
\newcommand{\kmpspMpc}{\hbox{$\kmps\Mpc^{-1}$}}
\newcommand{\civ}{\ion{C}{iv}}
\newcommand{\mgii}{\ion{Mg}{ii}}

\newcommand{\Cloudy}{{\small CLOUDY}}

\title[BLR line ratios: metallicity or density?]{High-ionization emission line ratios from quasar broad line regions: metallicity or density?}

\author[M. J. Temple et al.]{Matthew J. Temple,$^{1,2}$\thanks{E-mail: Matthew.Temple@mail.udp.cl}
Gary J. Ferland,$^{3}$
Amy L. Rankine,$^{2}$
Marios Chatzikos$^{3}$
and Paul C. Hewett$^{2}$
\\
$^{1}$N\'ucleo de Astronom\'ia, Universidad Diego Portales, Av. Ej\'ercito Libertador 441, Santiago, Chile\\
$^{2}$Institute of Astronomy, University of Cambridge, Madingley Road, Cambridge CB3 0HA, UK\\
$^{3}$Department of Physics and Astronomy, The University of Kentucky, Lexington, KY 40506, USA\\
}

\date{Accepted 2021 May 31. Received 2021 May 27; in original form 2021 March 22}

\pubyear{2021}

\begin{document}
\label{firstpage}
\pagerange{\pageref{firstpage}--\pageref{lastpage}}
\maketitle

\begin{abstract}
The flux ratios of high-ionization lines are commonly assumed to indicate the metallicity of the broad emission line region in luminous quasars.
When accounting for the variation in their kinematic profiles,
we show that the \ion{N}{V}/\ion{C}{IV}, (\ion{Si}{IV}+\ion{O}{IV}])/\ion{C}{IV} and  \ion{N}{V}/Ly\,$\alpha$  line ratios do not vary as a function of the quasar continuum luminosity, black hole mass, or accretion rate.
Using photoionization models from {\small CLOUDY}, we further show that the observed changes in these line ratios can be explained by emission from gas with solar abundances, if the physical conditions of the emitting gas are allowed to vary over a broad range of densities and ionizing fluxes.
The diversity of broad line emission in quasar spectra can be explained by a model with emission from two kinematically distinct regions, where the line ratios suggest that these regions have either very different metallicity or density. 
Both simplicity and current galaxy evolution models suggest that  near-solar abundances, with parts of the spectrum forming in high-density clouds, are more likely.
Within this paradigm, objects with stronger outflow signatures show stronger emission from gas which is denser and located closer to the ionizing source, at radii consistent with simulations of line-driven disc-winds.
Studies using broad-line ratios to infer chemical enrichment histories should consider changes in density and ionizing flux before estimating metallicities.

\end{abstract}




\section{Introduction}


Over the past few decades, it has been noted that the ratios of various high-ionization broad emission lines in quasar spectra are expected to change if the metallicity of the emitting gas is altered, assuming all other factors remain constant
\citep[e.g.][]{1992ApJ...391L..53H, 
1993ApJ...418...11H,
1999ARA&A..37..487H,
2006A&A...447..157N}.
Many authors have then assumed that the metallicity of the emitting gas is the \emph{only} factor which can significantly affect these line ratios, and thus have used high-ionization line ratios as metallicity indicators to quantify the chemical enrichment in quasar broad line regions (BLRs).
Such studies typically find that the metallicity of the BLR is super-solar, and increases with increasing quasar luminosity \citep{1993ApJ...418...11H, 2006A&A...447..157N}, with increasing black hole mass \citep{2018MNRAS.480..345X}, with increasing accretion rate \citep{2020arXiv200914177S} and with stronger emission-line outflow signatures \citep{2012ApJ...751L..23W, 2017ApJ...835...24S}.
At the same time, the line ratios are not observed to evolve as a function of redshift, with  high-redshift quasars, up to $z\simeq7.5$, displaying essentially indistinguishable line properties to those at $z\simeq2$ \citep{2020ApJ...898..105O}. 
Within this paradigm, these objects require metallicities of at least five times that of the Sun \citep{2009A&A...494L..25J}, suggesting rapid chemical enrichment in the nuclei of massive galaxies within $\approx500$\,Myr of the formation of the first stars, which is then sustained (without increasing the nuclear metallicity any further) for at least the next 3\,Gyr. 
Furthermore, this highly enriched nuclear material cannot form significant numbers of stars, as the stellar metallicities in massive galaxies are seen to decrease at high redshifts, and are not thought to be significantly above solar \citep{2017MNRAS.467..115D, 2021arXiv210211514B}.

However, while it is undoubtedly true that the metal content of the BLR gas will have an effect on the observed emission-line ratios, the line emission strengths will also depend on the physical conditions of the emitting gas.
For example, \citet{2017ApJ...834..203S} showed that the lower-ionization line ratio \ion{Mg}{II}/\ion{Fe}{II} was strongly affected by the gas cloud density, which needs to be accounted for before inferring the Mg/Fe abundance ratio (see also \citealt{2021ApJ...907...12S}).
In this paper we demonstrate that the full range of high-ionization broad-line ratios observed in high-luminosity, $L_\textrm{bol}\simeq 10^{46-47}$\,erg\,s$^{-1}$, quasar spectra with redshifts $z\simeq2$, can be explained by varying the physical conditions of the emitting gas, in particular the density and the flux of ionizing photons, without needing to invoke changes in metallicity.
When presenting the results of photoionization modelling, we therefore assume throughout this work that the emitting gas has metallicity equal to that of the Sun, although we note that our results would be qualitatively unchanged if we instead adopted a fixed metallicity of a few times solar.
To enable a direct comparison with \citet{2018MNRAS.480..345X}, we 
study the rest-frame ultraviolet
\ion{N}{V}/\ion{C}{IV} and (\ion{Si}{IV}+\ion{O}{IV}])/\ion{C}{IV}  line ratios.
We also want a clean diagnostic of the density of the emitting gas, and the {\ion{N}{V}/Ly\,$\alpha$} ratio is found to provide this
(Section {\ref{sec:models}}): it has a strong dependence on density and ionizing flux but is well-bounded across parameter space. Ly\,$\alpha$ is strong enough to allow the {\ion{N}{V}/Ly\,$\alpha$} ratio to be estimated despite blending, and the kinematics of Ly\,$\alpha$ are found to be consistent with those of {\ion{N}{V}} and {\civ} (Section ~{\ref{sec:data}}, see also \citealt{2006A&A...447..157N}).

We measure the
Ly\,$\alpha$, \ion{N}{V}, (\ion{Si}{IV}+\ion{O}{IV}]), and \ion{C}{IV}
line ratios in Section~\ref{sec:data}, present the results of our photoionization modelling in Section~\ref{sec:models}, and discuss the implications in Section~\ref{sec:discuss}.
A flat $\Lambda$CDM cosmology with $\Omega_m=0.27$, $\Omega_{\Lambda}=0.73$, and $\textrm{H}_0=71 \kmpspMpc$ is assumed. All emission lines are identified with their wavelengths in vacuum in units of \AA ngstr\"{o}ms.

\section{Observational Data}
\label{sec:data}

\subsection{Quasar Sample}
\begin{figure*}
	\includegraphics[width=2\columnwidth]{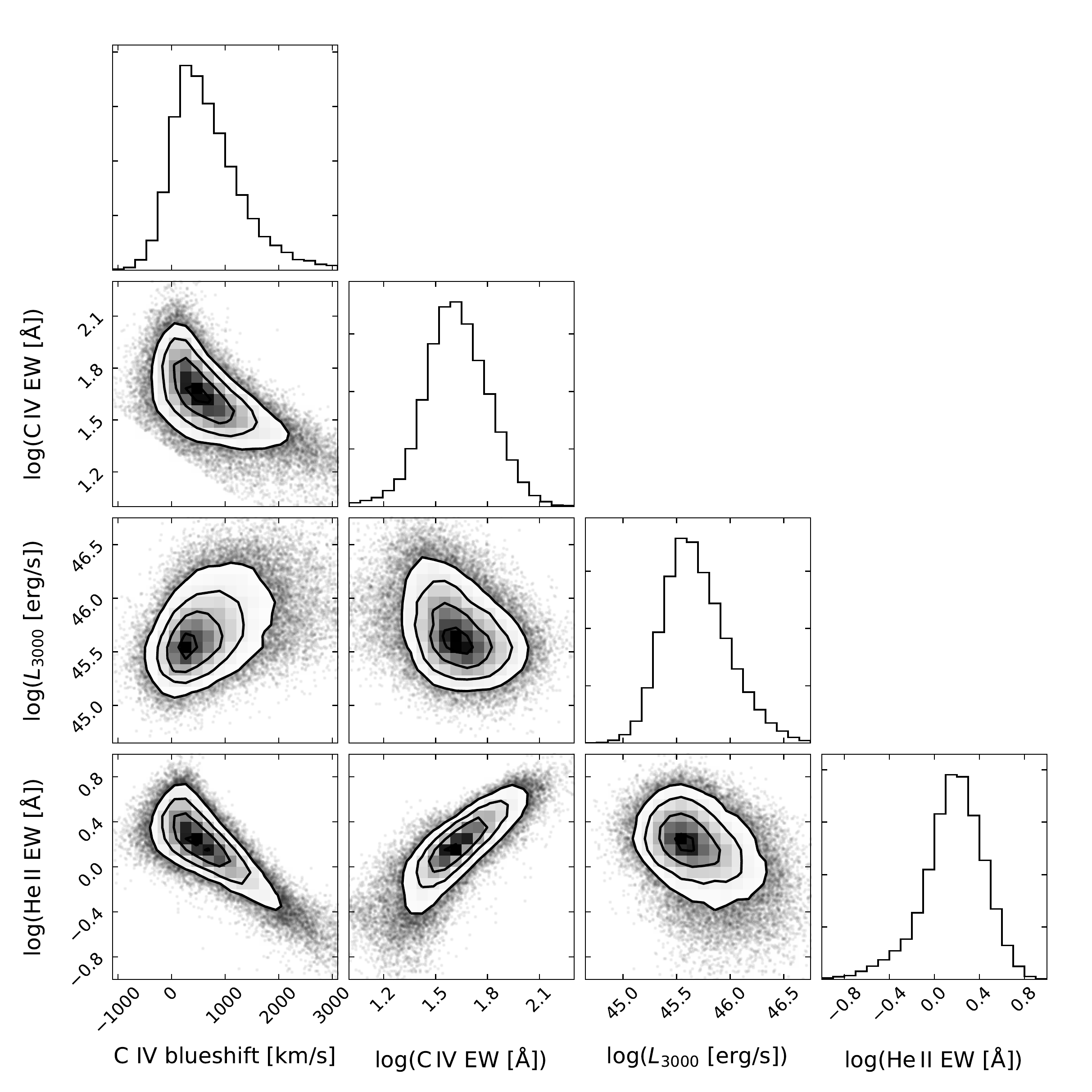}
    \caption{Key properties of our sample of 48\,588 quasars. Each object is selected to have redshift $2.0<z<3.5$ and to be free from \civ\ absorption signatures.
    The \civ\ blueshift, which quantifies the amount of emission from outflowing gas relative to the strength of emission at the systemic redshift,  correlates with the 3000\,\AA\ luminosity. The \civ\ EW anti-correlates with \civ\ blueshift, giving rise to a Baldwin effect. The strength of the \ion{He}{II}\,$\lambda$1640 recombination line is also known to correlate with the \civ-emission morphology.}
    \label{fig:sample1}
\end{figure*}

We take a sample of 48\,588 quasars with \civ\,$\lambda$1549 emission line measurements from \citet{Rankine20}. 
We take the redshift range  $2.0<z<3.5$, such that all objects have spectra from the Sloan Digital Sky Survey \citep[SDSS;][]{DR14Q} covering the rest-frame 1200-1700\,\AA\ region.
For our chosen redshift range, this parent sample consists of all quasars from the fourteenth data release of the SDSS
with average signal-to-noise per pixel $S/N>5$.

We exclude all objects with \civ\ absorption features (either broad or narrow) to avoid  emission-line flux measurements compromised by the presence of absorption.
\civ\ emission properties are taken from the \citet{Rankine20} catalogue, and rest-frame 3000\,\AA\ luminosities ($L_{3000}$) are determined by fitting a quasar SED model (Temple, Hewett \& Banerji in prep.) to the SDSS \textit{griz} photometry.
Systemic redshifts are calculated using the rest-frame 1600-3000\,{\AA}  region
\citep[section 3]{Rankine20},
to avoid biasing the  redshift determination in cases when \civ, \ion{N}{V} and Ly\,$\alpha$ are skewed with asymmetric emission profiles.

The \civ\ emission line `blueshift' is defined as the Doppler shift of the median line flux:
\begin{equation}
    \textrm{\civ\ blueshift} \equiv c\times (1549.48\,\AA-\lambda_\textrm{median})/1549.48\,\AA
\end{equation}
where $c$ is the speed of light and $\lambda_{\rm median}$ is the wavelength bisecting the total continuum-subtracted line flux. The wavelength 1549.48\,\AA \ is the mean wavelength of the \civ\,$\lambda\lambda$1548.19,1550.77 doublet assuming equal contributions to the emission.
This blueshift is a measure of the balance of emission from outflowing gas to emission from gas at the systemic redshift, and is known to anti-correlate with the equivalent width (EW) of \civ\
 \citep{2004ApJ...611..107L, Richards11, Rankine20}.
The \civ\ emission-line properties are also correlated with the ultraviolet continuum luminosity:
more luminous quasars  tend to show weaker \civ\ emission (the `Baldwin effect') which is more strongly blueshifted, indicative of stronger emission from ionized gas outflows and weaker emission from the symmetric `core' component of the line. These trends can also be seen in  Fig.~\ref{fig:sample1}.

\subsection{Black hole mass and Eddington fraction}
\label{sec:bhm}

\begin{figure*}
	\includegraphics[width=2\columnwidth]{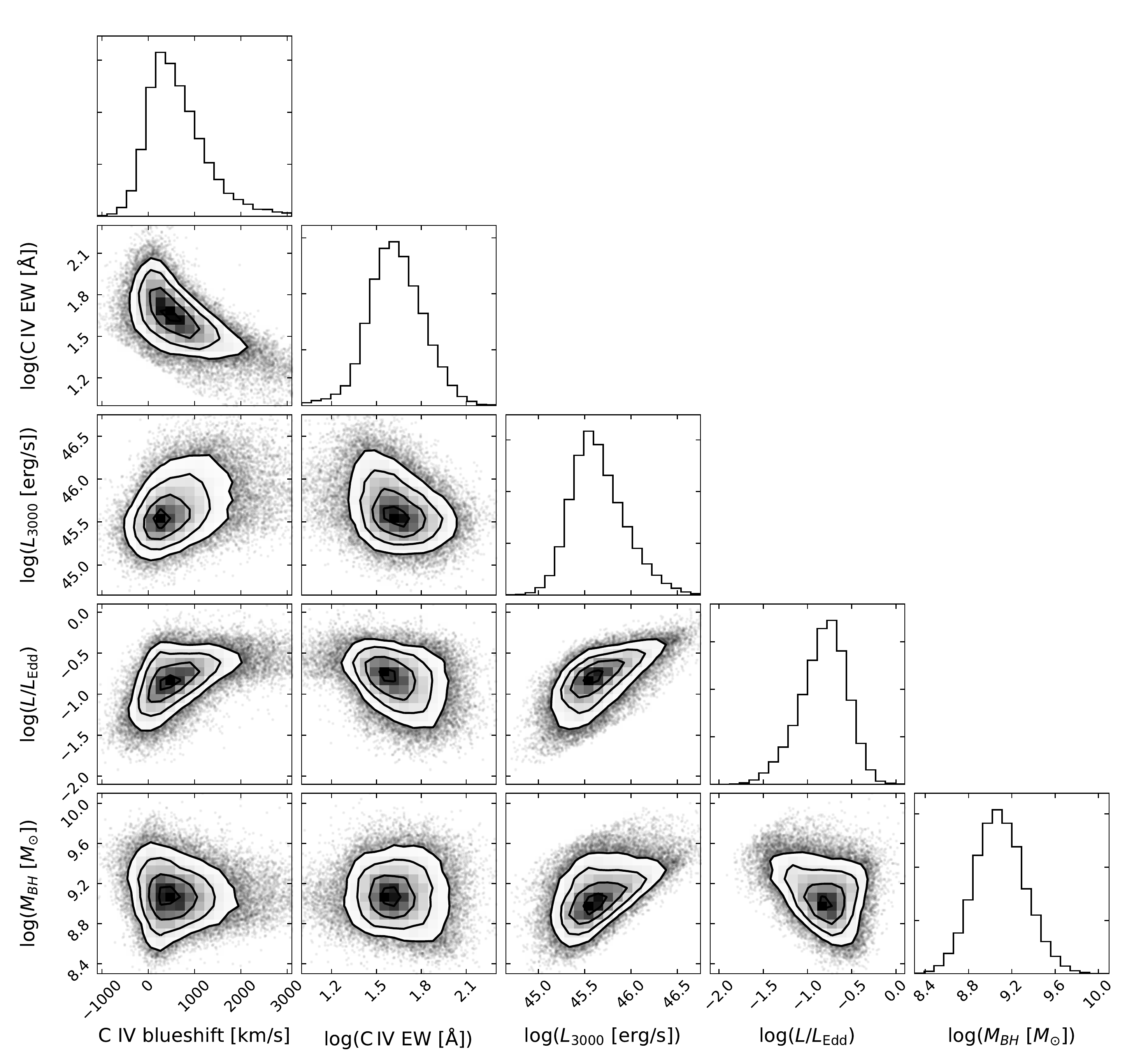}
    \caption{Black hole mass and Eddington-fraction estimates for 39\,833 objects from our sample which have coverage of \ion{Mg}{II}. Objects with strong wind signatures (large \civ\ blueshifts) are only found at higher Eddington fractions.}
    \label{fig:sample2}
\end{figure*}

Work by \citet{Coatman16, Coatman17} has shown that black hole mass estimates using the (uncorrected) velocity width of the \civ\ line can be biased as a function of line blueshift by up to an order of magnitude, i.e. a factor of 10 in the highest-blueshift objects.
When estimating black hole masses and Eddington fractions, we therefore restrict our sample to the 39\,833 objects with spectral coverage of \ion{Mg}{II}\,$\lambda$2800 {($2<z\lesssim2.7$)}.
In contrast to \citet{2018MNRAS.480..345X}, who used a \civ-based estimator, we estimate black hole masses using the \ion{Mg}{II}-based  estimator described by \citet{Vestergaard09}:
\begin{equation}
    M_\textrm{BH} = 10^{6.86}
    \left(\frac{\textrm{{\small FWHM}(\mgii)}}{1000\kmps}\right)^2
    \left(\frac{L_{3000}}{10^{44}\ergps}\right)^{0.5}M_\odot.
\end{equation}

Eddington ratios are calculated assuming a constant bolometric correction of $L_{\rm bol} = 5.15 \times L_{3000}$. However, we note that this correction may in fact be changing systematically as a function of \civ\ blueshift, as discussed by \citet{2021MNRAS.501.3061T}.
The properties of the 39\,833 object subsample are shown in Fig.~\ref{fig:sample2}.

\subsection{Method}

\begin{figure*}
	\includegraphics[width=1.94\columnwidth, clip=on, trim={0 5 0 22}]{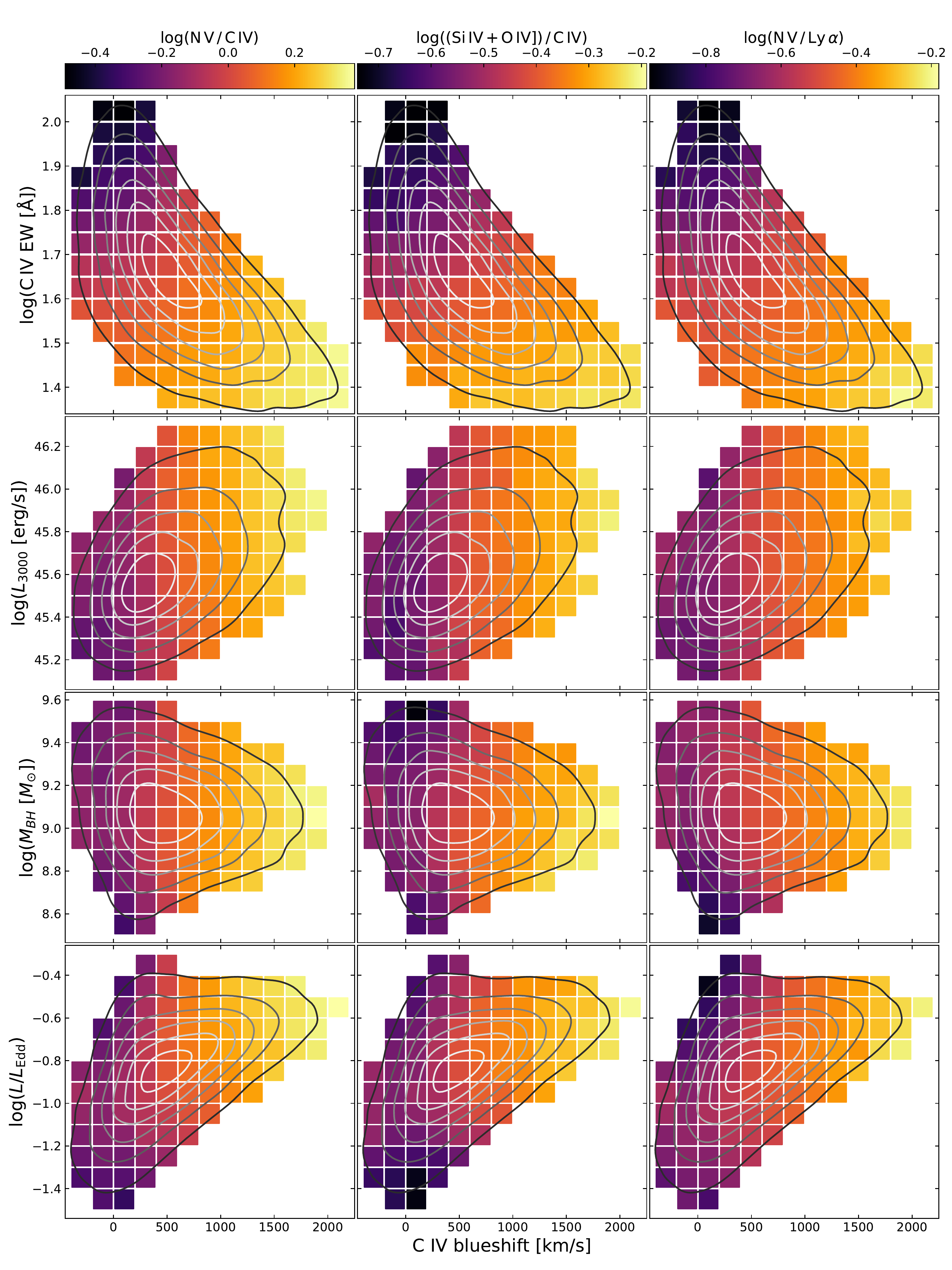}
    \caption{
    \textit{Left-right:}
    The observed \ion{N}{V}/\ion{C}{IV}, (\ion{Si}{IV}+\ion{O}{IV}])/\ion{C}{IV} and   \ion{N}{V}/Ly\,$\alpha$ line ratios in composite spectra taken from different regions of parameter space; each region is shown only if it contains at least 100 objects.
    Contours show the number density of quasars in evenly-spaced linear intervals: each projection of parameter space is seen to be unimodal.
    \textit{Top:}
    The line ratios change smoothly across the \civ\ blueshift-EW space.
    \textit{Below:} 
    When considering objects with the same \civ\ blueshift, the high-ionization line ratios do not change as a function of luminosity, black hole mass, or Eddington fraction. However, brighter objects and higher accretion-rate objects are more likely to display stronger outflow signatures, and thus are more likely to show stronger \ion{N}{V}, weaker \civ\ and weaker Ly\,$\alpha$.
    An interactive version of this plot is available online. 
    }
    \label{fig:results}
\end{figure*}

\begin{figure*}
	\includegraphics[width=1.89\columnwidth]{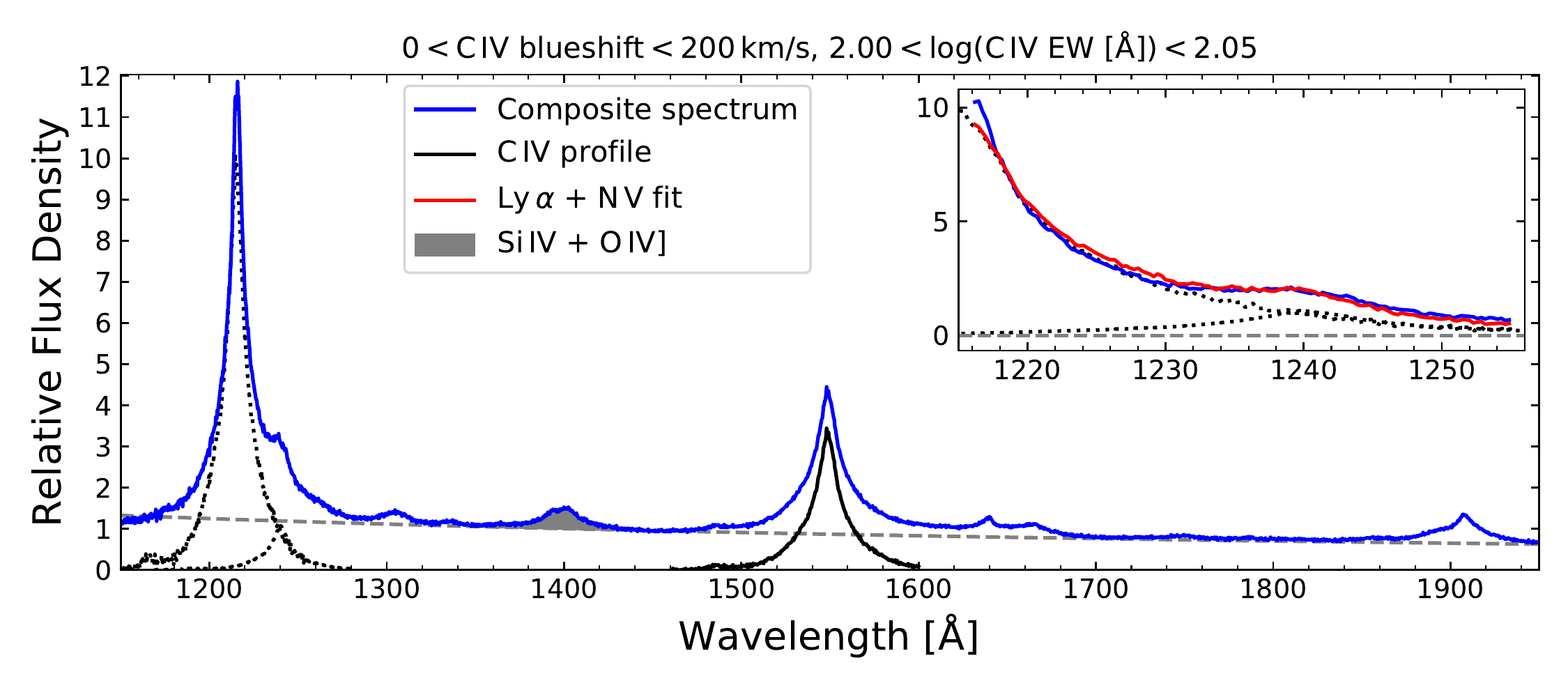}
	\includegraphics[width=1.89\columnwidth]{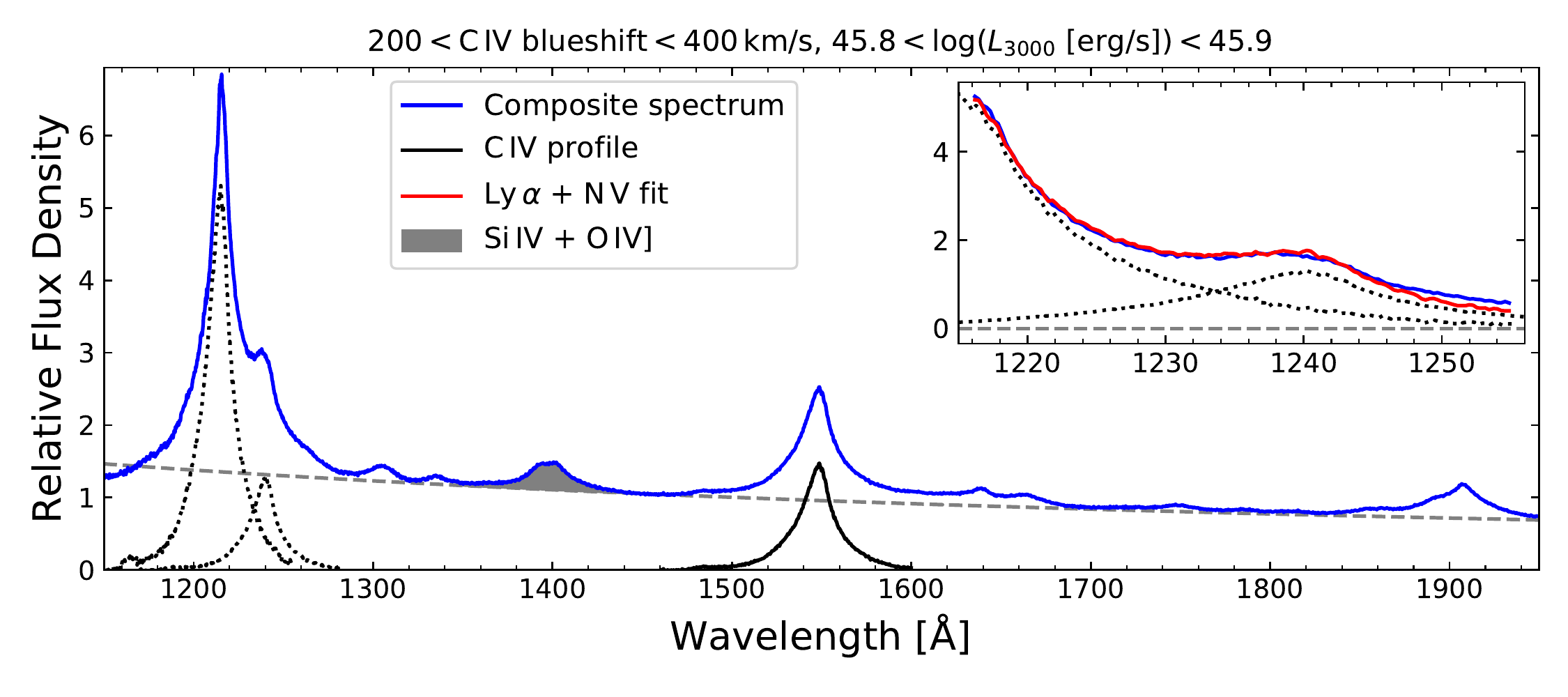}
	\includegraphics[width=1.89\columnwidth]{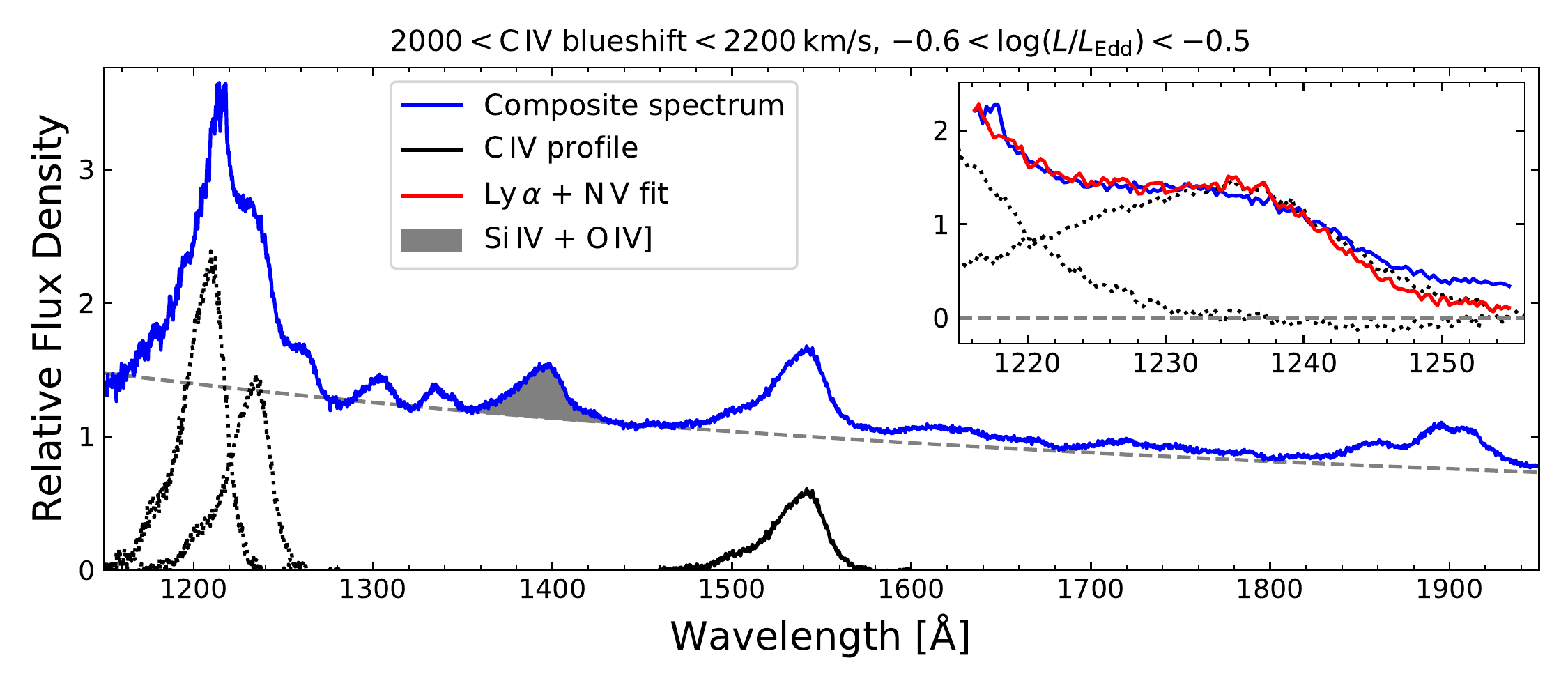}
    \caption{Three examples of our method to measure emission line ratios in composite spectra.
    Before forming the composite, each individual spectrum is normalised relative to the average flux density across 1300-2200{\,\AA}.
     \textit{Top:}  `core-dominated' composite with strong, symmetric emission lines.
     \textit{Middle:}  from  the more densely-populated central region of parameter space, more representative of a typical $z\simeq2$ quasar.
     \textit{Bottom:}  `wind-dominated' composite with large \civ\ blueshift and weak systemic emission.
    \textit{Inset:} the continuum-subtracted 1216-1255\,\AA\ wavelength region used to quantify Ly\,$\alpha$ and \ion{N}{V}, which are both constrained to have the same kinematic profile as \civ.
    The flux at wavelengths shortwards of 1215\,\AA\ is suppressed due to absorption from neutral hydrogen in the inter-galactic medium, and so we do not fit to the blue wing of Ly\,$\alpha$. The excess of flux at 1255\,\AA\ is due to emission from \ion{Si}{II}\,$\lambda$1263.
    }
    \label{fig:fits}
\end{figure*}

Measuring the strengths of the Ly\,$\alpha\,\lambda$1216, \ion{N}{V}$\,\lambda$1240, (\ion{Si}{IV}+\ion{O}{IV}])$\,\lambda$1400 and \ion{C}{IV}$\,\lambda$1549 emission lines,
we wish to investigate how the
\ion{N}{V}/\ion{C}{IV}, (\ion{Si}{IV}+\ion{O}{IV}])/\ion{C}{IV} and \ion{N}{V}/Ly\,$\alpha$ line ratios
change 
across the quasar population.
As shown in Fig.~\ref{fig:results}, we consider four projections of the five-dimensional parameter space 
defined by \civ\ blueshift, luminosity, black hole mass, Eddington fraction and \civ\ EW.
To determine the systematic behaviour of the line ratios accurately, each projection of the quasar parameter space is divided into bins and 
high signal-to-noise ratio (S/N)
composite spectra are constructed using the quasars in each bin. 
Before stacking, each  spectrum is normalised relative to the average flux across the 1300-2200\,{\AA} wavelength range. We take the median of each stack to form the composite, although we have verified that our results would remain unchanged if we instead took the mean.

All bins are chosen with \civ\ blueshift in 200\kmps\ intervals. 
We explore the dependence of the line ratios on luminosity, black hole mass, Eddington fraction and \civ\ emission-line strength separately,
with \civ\ EW, $L_{3000}$, $M_\textrm{BH}$ and $L/L_\textrm{Edd}$ binned in intervals of 0.05, 0.1, 0.1 and 0.1\,dex respectively.
We only consider bins which contain at least 100 quasars.

In each binned region of projected parameter space,
a composite of the SDSS spectra in that region is formed,
and line ratios are measured as follows.
The \civ\ line profile is isolated by subtracting a power-law continuum defined using the flux in 10\,\AA \ windows centred at 1450 and 1625\,\AA,
which avoids biasing the shape of the \civ\ profile with flux from the shelf of emission at $\approx$1600-1660\,\AA. 
The \civ\ line flux is determined by integrating the line profile over the 1500-1600\,\AA\ wavelength interval.
A second power-law continuum is defined using the flux in 10\,\AA \ windows centred at 1450 and 1978\,\AA. The resulting power-law is subtracted from the 1216-1450\,\AA\ region.
The continuum-subtracted flux between 1216-1255\,\AA\ is then fit by adopting the \civ\ line profile for both the Ly\,$\alpha$ and \ion{N}{V} lines.
The blue wing of Ly\,$\alpha$ is not included in the fitting window as we have not corrected for Lyman suppression in the inter-galactic medium, and $\lambda<1255$ is chosen to minimise contamination from the low-ionization \ion{Si}{II} line at 1263\,\AA.
By constraining the Ly\,$\alpha$ and \ion{N}{V} lines to have the same shape as \civ\ we are effectively following the same prescription as \citet{2006A&A...447..157N}. The Ly\,$\alpha$+\ion{N}{V} complex is found to be well-fit across the full range of \civ\ morphologies when using this method.
The \ion{Si}{IV}+\ion{O}{IV}] flux is determined by integrating each continuum-subtracted composite over the wavelength interval 1350-1450\,\AA.
We show three examples of the line-fitting routine 
in Fig.~\ref{fig:fits};
the equivalent plot for each composite is made available online.

\subsection{Results}

We show the derived \ion{N}{V}/\ion{C}{IV}, (\ion{Si}{IV}+\ion{O}{IV}])/\ion{C}{IV} and \ion{N}{V}/Ly\,$\alpha$ line ratios in Fig.~\ref{fig:results}. The contours show the distribution of quasars across each parameter space.
Our method to derive line ratios is very similar to that of \citet{2006A&A...447..157N},
and the range of ratios we observe is essentially the same as found by previous authors \citep[e.g.][]{2006A&A...447..157N, 2012ApJ...751L..23W}.

From Fig.~\ref{fig:results}, we can see immediately that \civ\ morphology is strongly correlated with the high-ionization line ratios, similar to the trend seen in the equivalent width of \ion{He}{II}\,$\lambda$1640 \citep[][fig. 12]{Rankine20}. Objects with large \civ\ blueshifts tend to have stronger \ion{N}{V} and (\ion{Si}{IV}+\ion{O}{IV}]), and weaker Ly\,$\alpha$ and \ion{He}{II}. Conversely, quasars with high-EW \civ\ also tend to have strong Ly\,$\alpha$ and \ion{He}{II}, and weaker \ion{N}{V} and (\ion{Si}{IV}+\ion{O}{IV}])  emission.

From Fig.~\ref{fig:results}, we can also see that, when controlling for the \civ\ morphology, there is no significant change in the \ion{N}{V}/\ion{C}{IV}, (\ion{Si}{IV}+\ion{O}{IV}])/\ion{C}{IV} and   \ion{N}{V}/Ly\,$\alpha$ line ratios with varying luminosity, black hole mass or Eddington fraction.
However, more luminous quasars will, on average, display larger \civ\ blueshifts (i.e. brighter objects are more likely to show evidence for outflows) and so more luminous quasars will tend to show enhanced \ion{N}{V} and  (\ion{Si}{IV}+\ion{O}{IV}]) and weaker \civ\ and Ly\,$\alpha$.
The trends we observe are therefore consistent with the correlations between luminosity and line ratios reported by \citet{2006A&A...447..157N},
and with the correlations between line kinematics and line ratios found by 
\citet{2012ApJ...751L..23W} and \citet{2017ApJ...835...24S}.

Our results are inconsistent with those of \citet{2018MNRAS.480..345X}, who found evidence for correlations between the black hole mass and both the \ion{N}{V}/\ion{C}{IV} and (\ion{Si}{IV}+\ion{O}{IV}])/\ion{C}{IV} line ratios.
However, \citet{2018MNRAS.480..345X} estimated black hole masses using a \civ-based estimator. 
As noted in Section~\ref{sec:bhm}, such estimators are biased (when uncorrected) as a function of \civ\ blueshift, and our black hole mass estimates are less biased by virtue of the fact they are derived using the \ion{Mg}{II} emission line.
We therefore suggest that the correlations found by \citet{2018MNRAS.480..345X} are spurious,
in the sense that they can be attributed to objects in their sample with increasing \civ\ blueshifts
(which  have different high-ionization line ratios)
possessing black hole mass estimates which are increasingly biased.

\subsection{
Line ratios from `wind-dominated' and `core-dominated' composite spectra
}

\begin{figure*}
	\includegraphics[width=2\columnwidth]{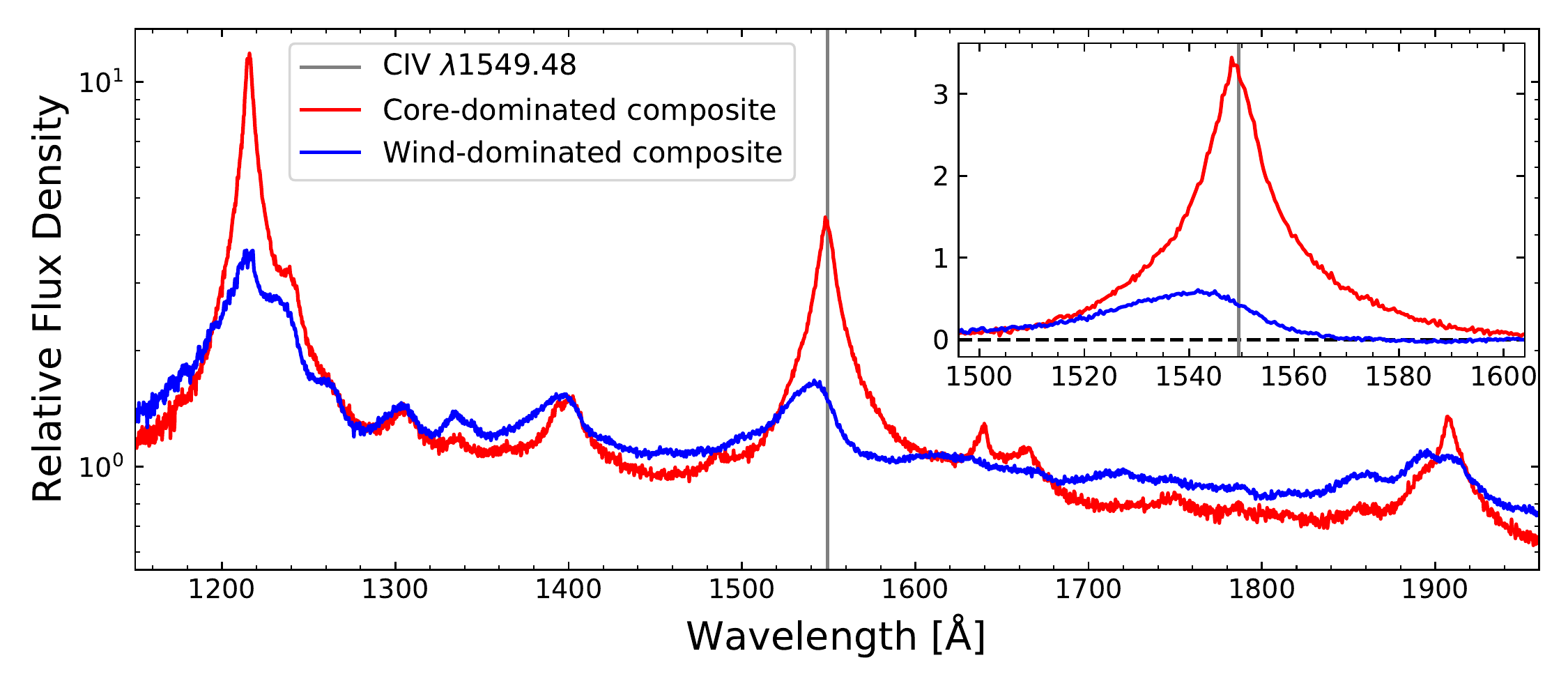}
    \caption{{Comparison of `wind-dominated' and `core-dominated' composite spectra from Fig.~{\ref{fig:fits}}, with the rest wavelength of {\civ} marked in gray.
    Note the logarithmic scale on the y-axis.
    \textit{Inset:} the continuum-subtracted {\civ} profiles in each composite. 
     The `core-dominated' {\civ} profile is symmetric with no excess in the blue wing of the line, while our `wind-dominated' {\civ} profile is almost entirely due to emission from outflowing material. }
    }
    \label{fig:WindCore}
\end{figure*}

The smooth changes in high-ionization
line ratios seen as a function of \civ\ morphology are consistent with a model in which the high-ionization line emission is due to two kinematically distinct components: one at the systemic redshift of the quasar and one which is blueshifted, hereafter referred to as the `core' and `wind' respectively.
The results presented in Fig.~\ref{fig:results} show that these two components have significantly different line ratios, suggesting that they arise from regions with different physical conditions.

We would like to isolate these two components, in order to compare their observed line ratios to those predicted by photoionization models. The binned regions of parameter space shown in the top and bottom panels of Fig.~\ref{fig:fits} are typical of  the extrema of the observed range  of \civ\ morphology. 
We compare these composites in Fig.~{\ref{fig:WindCore}}.
While we cannot rule out the possibility they still contain some small fraction of emission from the other component, 
the {\civ} profile in the `core-dominated' composite is entirely symmetric with no excess in the blue, while the {\civ} profile in the `wind-dominated' objects contains very little emission redwards of 1549.48\,{\AA}. Any attempt to further decompose the line emission in these composites would be somewhat degenerate, and thus would not improve our constraints on which regions of parameter space could plausibly be giving rise to the observed line emission in each `component'.
We therefore take the line ratios from these two regions
with extreme {\civ} profiles
to compare with the results of photoionization models.
This yields
(\ion{N}{V}/\ion{C}{IV}, (\ion{Si}{IV}+\ion{O}{IV}])/\ion{C}{IV},   \ion{N}{V}/Ly\,$\alpha$)
line ratios of (0.32, 0.18, 0.11) 
and (2.3, 0.59, 0.64) respectively
in our `core-dominated' and `wind-dominated' composite spectra.

We note in passing that the aim of this work is to provide a plausible scenario (which need not invoke large changes in metallicity) to explain the observed line ratios in the high-ionization ultraviolet lines, which show strong signatures of outflows from the BLR. We therefore do not attempt to constrain the properties of lower-ionization lines, which display very different kinematics, and are therefore most likely arising from different locations in physical space.

\section{Photoionization modelling}
\label{sec:models}

\subsection{Numerical calculations}

\begin{figure*}
	\includegraphics[width=2\columnwidth]{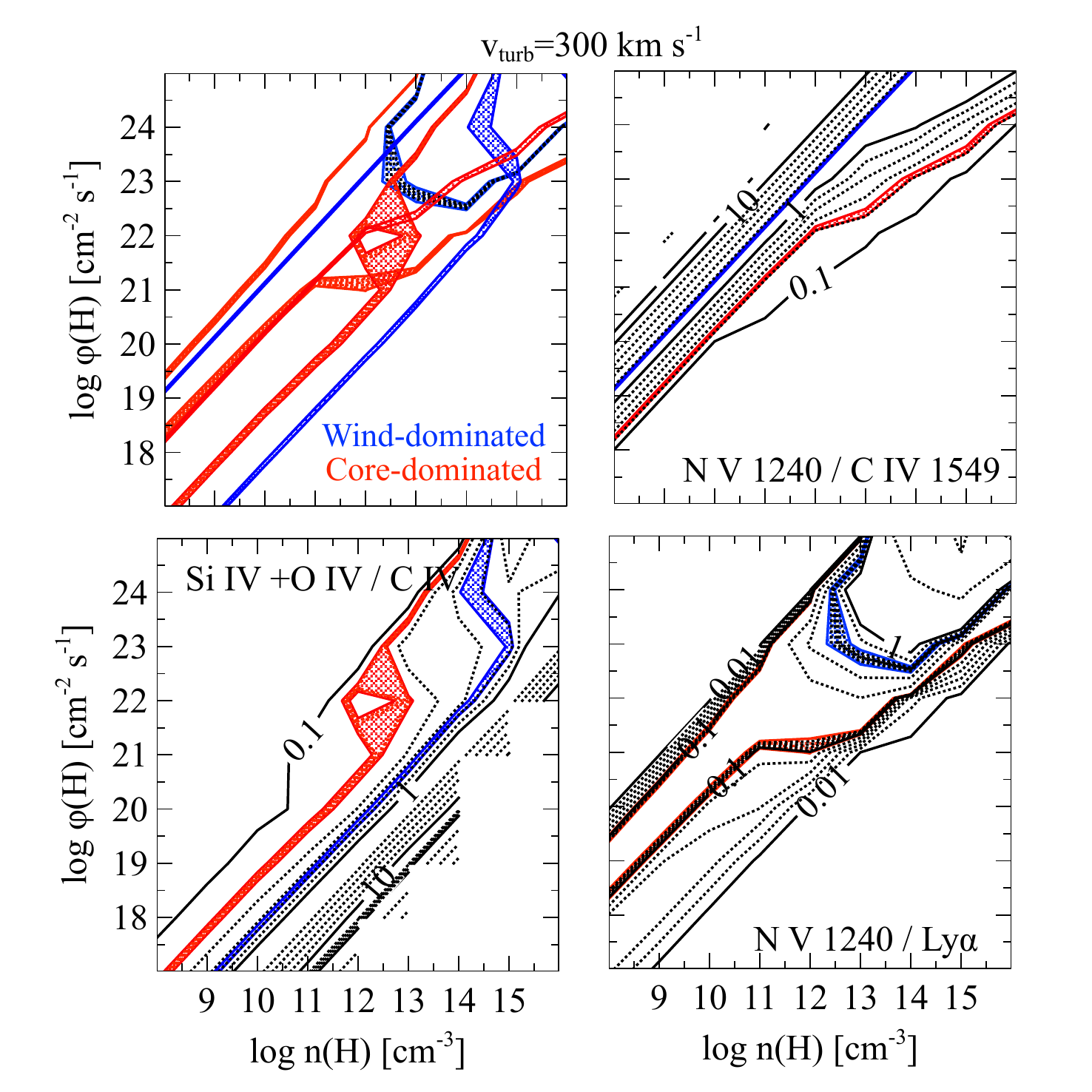}
    \caption{
    \textit{Black contours:}
    photoionization model predictions for the line ratios shown in Fig~\ref{fig:results}, assuming solar metallicity. Contours in
    \text{blue}
    give the line ratios seen in objects where the high-ionization lines are dominated by emission from a `wind' component;
    those in 
    \text{red} correspond to
    line ratios seen in objects which are dominated by emission from a more symmetric `core' component at the systemic redshift.
    The \ion{N}{V}/\ion{C}{IV} and (\ion{Si}{IV}+\ion{O}{IV}])/\ion{C}{IV} ratios are mostly sensitive  to just the ionization parameter, with plausible solutions predicted across a wide range of $n_\textrm{H}$ and $\phi$.
    \textit{Bottom right:} the \ion{N}{V}/Ly\,$\alpha$ ratio has more diagnostic power, constraining the wind-dominated emission to high densities and ionizing fluxes.
    \textit{Top left:}
    combining the three line ratios, we see that, modulo some contamination from gas in other regions of parameter space, the line ratios observed in wind-dominated spectra could plausibly be explained by emission from a high-$n_\textrm{H}$, high-$\phi$ region of parameter space, without needing to change the metallicity of the emitting gas.}
    \label{fig:theory6}
\end{figure*}

\begin{figure}
	\includegraphics[width=1\columnwidth]{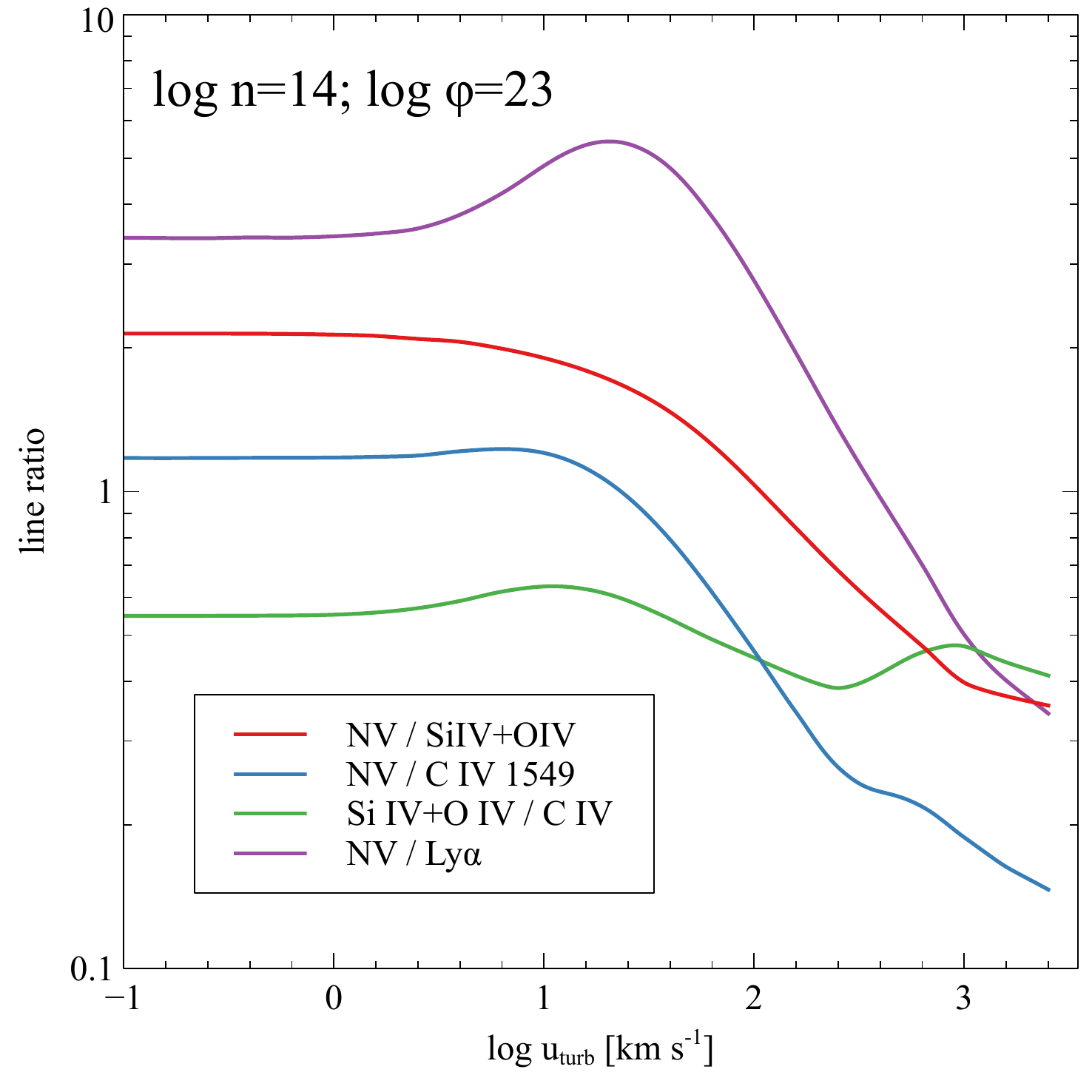}
    \caption{
    Ratios of the \ion{N}{V} doublet to other strong lines as a function of turbulent velocity.
    Calculations are for the indicated density and flux.  The lines are
    thermalized so the line ratios are near the blackbody limit.
    Line optical depths decrease as the turbulence increases so clouds with the smallest
    turbulence have the largest optical depths, producing lines closer to the
    blackbody limit.}
    \label{fig:v}
\end{figure}

In Fig.~\ref{fig:theory6}, we show the
\ion{N}{V}/\ion{C}{IV}, (\ion{Si}{IV}+\ion{O}{IV}])/\ion{C}{IV} and \ion{N}{V}/Ly\,$\alpha$
line ratios predicted by photoionization modelling.
The model set-up is the same as described in \citet{2020MNRAS.496.2565T}, 
using version 17.03 of \Cloudy\ \citep{2017RMxAA..53..385F}. 
The adopted SED is the intermediate $L/L_\textrm{Edd}$ case
described by \citet{2020MNRAS.494.5917F}.
To be consistent with our previous models we
assume a cloud column density of $10^{23}$\,cm$^{-2}$ and use the complete
models of \ion{Fe}{III} emission adopted by \citet{2020MNRAS.496.2565T}
and \ion{Fe}{II} described by \citet{ 2021ApJ...907...12S}.
These data sets are included in the C17.03 download.

We present predictions for a broad range of cloud densities and
flux of ionizing photons $\phi$(H).  
We assume solar metallicities and a fixed microturbulence parameter of 300\kmps.
The microturbulence introduces a term in the Voigt function that
account for line broadening over scales shorter than a photon
mean free path \citep{2014tsa..book.....H}.
Moderate (100-300\kmps) microturbulence has been shown to be necessary to reproduce the observed properties of \ion{Fe}{III}  and \ion{Fe}{II} ultraviolet line emission \citep{2020MNRAS.496.2565T, 2021ApJ...907...12S}.
The effect of varying this parameter is shown in Fig.~\ref{fig:v}.

The $x$-axis of Fig.~\ref{fig:theory6} gives the cloud density $n_\textrm{H}$. The $y$-axis gives the flux of ionizing photons, $\phi({\rm H})$, which scales with the  distance from the ionizing source as $\phi \propto r^{-2}$ (assuming the source of ionizing photons is point-like). Larger values on the y-axis therefore correspond to emission from closer in to the central black hole. 
We prefer this parameter over the more commonly used ionization
parameter $U$ (see equation \ref{eq:IonizationParameter})
since it is related to
distance from the ionizing source,
which can be measured with reverberation studies. 
The predicted line ratios across the parameter space are shown by contours. The \civ\ emission is known to peak along a diagonal corresponding to constant values of the ionization parameter
\begin{equation}
\label{eq:IonizationParameter}
U=\frac{\phi({\rm H})}{c\, n_\textrm{H}}.
\end{equation}
Predictions for  \ion{N}{V}/\ion{C}{IV} and (\ion{Si}{IV}+\ion{O}{IV}])/\ion{C}{IV}
are therefore not given in the top-left and bottom-right corners of the flux-density parameter space as the \civ\ strength in the denominator tends to zero.

\subsection{Photoionization modelling results}

The \ion{N}{V}/\ion{C}{IV} and (\ion{Si}{IV}+\ion{O}{IV}])/\ion{C}{IV} ratios shown in Fig.~\ref{fig:theory6} are seen to largely depend on the ionization parameter, with plausible solutions predicted across the full range of density $n_\textrm{H}$ and ionizing flux $\phi$.
However, the \ion{N}{V}/Ly\,$\alpha$ ratio has more diagnostic power,
with large values of this ratio only predicted from 
regions of high density and high ionizing flux.

To compare the results of our photoionization modelling to the range of line ratios observed across the quasar population, we colour contours representing  the extreme line ratios seen in our `wind-dominated' and `core-dominated' composites in blue and red respectively.
The width of each coloured region has been set at 10 per cent of the measured line ratio for visualisation purposes.
In the top-left panel of Fig.~\ref{fig:theory6}, we combine these coloured contours.

We note that in a  model where the high-ionization line emission is coming from two distinct regions, corresponding to the `core' and `wind'  parts of the line, the `wind-dominated' and `core-dominated' composites which we have constructed might still be expected to show some emission from both regions, even if they are dominated by emission from a single region.
With this in mind, the `wind-dominated' line ratios seen in objects with high-blueshift \civ\ (and \ion{N}{V} and Ly\,$\alpha$) lines can be explained as arising mostly from gas with large $\phi({\rm H})$ and $n_\textrm{H}$,
while the objects with strong, symmetric \civ\ can then be explained as arising mostly from gas at a range of lower densities and fluxes, with emission explained using a broad range of cloud properties,
as in the  `locally optimally-emitting cloud' model \citep{1995ApJ...455L.119B}.
The physics responsible for the difference in line ratios predicted at high densities is discussed in Appendix~\ref{sec:bblim}.

We note that there are other physical parameters, aside from the metallicity, which could also be tuned to match the observations. In Fig.~\ref{fig:v}, we show the effect of varying the microturbulence parameter $\nu_\textrm{turb}$ on the measured line ratios for the dense, close-in point of parameter space we have identified with  the `wind' component. $\nu_\textrm{turb}$ accounts for Doppler broadening across scales shorter than the mean free path of a photon, which could correspond to turbulent motions or ordered motions as expected in a wind or in the accretion disc.
In particular, given the large velocity gradients which must be present in the outflowing wind, thermal line widths are clearly inappropriate for this component. We suggest that the line broadening in high-ionization BLR winds could plausibly range from $10^{2-3}$\,\kmps; here we have chosen to adopt $\nu_\textrm{turb}=300\,\kmps\approx10^{2.5}\,\kmps$ to maintain consistency with
\citet{2021MNRAS.501.3061T}.
A full discussion of the effect of varying $\nu_\textrm{turb}$ is beyond the scope of this paper; here we simply aim to show that there exist plausible solutions which do not require the metallicity of the BLR to vary.

Our analysis does not prove that the metallicity of the line emitting gas is solar, or close to solar.
Indeed, studies of individual AGN which, by selection, show relatively strong absorption features suggest that the metallicity of the outflowing absorbing material is around twice solar \citep[e.g.][]{2006ApJ...646..742G, 2007ApJ...658..829A, 2020A&A...633A..61A},
and our results are not inconsistent with their findings.
However, adopting a model with emission from two distinct (but chemically identical) components means that it is still possible to reproduce the systematic trends in line ratios as a function of key physical parameters, such as luminosity or Eddington ratio, that hitherto have been interpreted as due to significant variations ($\sim$1\,dex) in metallicity.
Compared to previous studies, the key differences in our two-component model are the inclusion of microturbulence ($\simeq$300\kmps), as expected in the case when lines are  broadened by more than just thermal pressure, 
and the higher gas density in the outflowing component, $n_\textrm{H}\simeq 10^{13-14}$\,cm$^{-3}$, corresponding to emission from closer to the ionizing source.

\section{Discussion}
\label{sec:discuss}

Measuring the chemical evolution of galaxies is key to understanding how they have formed and evolved over cosmic time.
Early `closed box' models of galactic evolution reached very high metallicity $Z$.
\citet{1992ApJ...391L..53H, 1993ApJ...418...11H, 1999ARA&A..37..487H}
showed that quasar spectra were consistent with $Z\approx 10 Z_\odot$.
Current galactic evolution models, which include a more sophisticated treatment of the infall of pristine gas, do not reach such large $Z$. 
In fact,  recent models of galaxy evolution suggest that the metallicity of the most massive galaxies (such as are expected to host luminous quasars) at redshift $z\simeq2$ is approximately solar, and the metallicity of low-mass  galaxies is sub-solar by at least an order of magnitude  \citep[e.g.][fig.~4]{2017MNRAS.467..115D}.
These models are motivated by, and in are agreement with, recent observations (\citealt{2019A&ARv..27....3M}, section 5, and references therein; \citealt{2020arXiv200907292S}, fig.~7; \citealt{2021arXiv210211514B}, fig.~6).
Observations also suggest that galaxy metallicity gradients are flatter at higher redshifts, suggesting that the interstellar medium in galaxy nuclei is not significantly over-enriched (\citealt{2019A&ARv..27....3M}, section 6; \citealt{2020MNRAS.492..821C}).

The large number of quasar spectra which are now available from the SDSS allows a more nuanced approach to analysing trends in the quasar population.
There are two distinct components to the high-ionization ultraviolet emission lines: a symmetric `core' that is consistent with the classic BLR, and an outflowing `wind' component which is responsible for the blueshifted emission.
The variation in the equivalent width of the outflow component from object to object is
approximately a factor of two, whereas the variation in the
systemic component is a factor of six or seven.
While there are a
small number of objects where one of the components is almost absent,
the vast majority of $z\simeq2$ quasar spectra
include significant emission from both systemic and outflow components.
These components should be deblended before attempting to infer the physical properties of the emitting media
in individual objects.

The standard emission line analysis, at face value, suggests very different metallicity for these two components, with a solar BLR and high-$Z$ wind \citep[e.g.][]{2020arXiv200914177S}. 
We note that it is still very possible that quasar BLRs have a modest range of metallicities, and that metallicity is one of the factors that produces favourable conditions for outflows to launch due to increased ultraviolet line opacities.
However, we have shown in this work that both components are consistent with solar metallicity if, as suggested by other emission line diagnostics, the wind has a very high density and ionizing photon flux \citep{2014ApJ...793..100M, 2020MNRAS.496.2565T}.
In other words, the population of SDSS quasar spectra is consistent with line emission from regions with a very broad range of density and with the near-solar abundances suggested by current galactic evolution theory.

There is a strong systematic dependence on the ratio of the systemic
and outflow components as a function of both luminosity and Eddington fraction. The
outflow component contribution increases with increasing luminosity -
the typical \civ\ blueshift is $\simeq$0\kmps\ at $10^{45}\ergps$ and $\simeq$2000\kmps\ at $10^{47}\ergps$, although there is  a large dispersion in blueshift at
fixed luminosity. An extrapolation of this relation to higher
luminosities is consistent with the properties of very high-redshift ($z>6$)
quasars, which (in current samples) often possess high luminosities, large line shifts, and weak Ly\,$\alpha$ emission.
The \civ\ blueshift has been shown by e.g. \citet{2004ApJ...617..171B} and \citet{2007ApJ...666..757S} to correlate with `Eigenvector 1' \citep{BG92} and, as such, our high-blueshift, wind-dominated objects are analagous to `Extreme Population A' quasars \citep{2000ApJ...536L...5S}.
Our results are therefore consistent with those of \citet{2020arXiv200914177S}, to the extent that we find distinct physical regions for `core' and `blueshifted' emission components. 
However, these regions  do not necessarily need to have extreme ($Z\approx50Z_\odot$) metallicities in order to explain the observed line ratios.

\subsection{Implications of fixed metallicity}

In Section~\ref{sec:models}, we demonstrated that if the metallicity of broad line regions does not change significantly across the quasar population, it is still possible to explain the diversity of high-ionization ultraviolet emission-line ratios by allowing for emission from two components of differing densities and illuminated by differing ionizing fluxes.

In particular, we have shown that emission from dense gas can explain the Ly\,$\alpha$ : \ion{N}{V} : \civ\  emission line ratios seen in objects with strong outflow signatures.
As one moves to higher densities, the lines move towards thermalization at the blackbody limit for their respective wavelengths. At the same time, the preponderance of \civ\ to emit at fixed ionization parameter, $U$, causes emission from large values of $\phi$ to be preferentially observed, corresponding to emission close-in to the ionizing source, which may or may not be the same source responsible for driving the wind.
In particular, we note that if the wind is  located this close to the central source, then it would still be possible for there to be low-density gas in the outflow, however at these radii one would not observe significant line emission: the value of $U$ would be such that   low-density gas would not emit {\civ}.

Hydrodynamic simulations of disc-wind models \citep{1995ApJ...451..498M, 1996ApJ...466..704C} have often built on the work of \citet{2000ApJ...543..686P} and \citet{2004ApJ...616..688P}.
These works considered a $10^8$\Msun\ black hole, emitting at 50 per cent of its Eddington limit, i.e., at $10^{46.7}$\ergps. Assuming the SED with $L/L_\textrm{Edd}=10^{-0.55}$ from \citet{2012MNRAS.425..907J}, this corresponds to an ionizing photon luminosity of $Q(\textrm{H})\simeq 4\times10^{56}$\,s$^{-1}$.
The simulations of \citet{2000ApJ...543..686P} and \citet{2004ApJ...616..688P} showed that such a system was capable of launching a wind through line driving from a radius of $10^{16}$\,cm, which corresponds to an ionizing flux of $\phi(\textrm{H}) = Q(\textrm{H})/(4\pi r^2) = 10^{23.5}$\,cm$^{-2}$\,s$^{-1}$.
We therefore suggest that the outflowing component which we observe in \civ, \ion{N}{V} and Ly\,$\alpha$,  in quasars with $L/L_\textrm{Edd}>0.25$, could be associated with this line-driven wind.
At these radii, the Newtonian escape velocity is $\sim$100\,000\,\kmps, larger than the average observed line shift, and so the wind in most objects would be expected to fail to escape the sphere-of-influence of the central supermassive black hole.

For such close-in and highly-illuminated gas, the densities at which we would preferentially observe \civ\ are around 3\,dex higher than those seen in the simulations of \citet{2004ApJ...616..688P}.
However,  recent work
 \citep[][]{2020ApJ...893L..34D, 2020MNRAS.492.5540M}
has suggested that
`clumpy' gas is needed to produce the physical conditions required to observe outflows in the high-ionization line profiles, 
which may help to reduce this tension.
More detailed investigations into simulations \citep{2010MNRAS.408.1396S, 2014ApJ...789...19H, 2020MNRAS.494.3616N} have highlighted challenges in launching disc-winds through line driving: further work is needed on both photoionization models and radiation-(magneto-)hydrodynamic simulations in order to reconcile observations of emission lines with a self-consistent model of quasar outflows.

We note in passing that the dense gas which constitutes the close-in wind component might also be expected to emit the iron K$\alpha$ line, as discussed  by \citet{2020ApJ...898..141D, 2021ApJ...906...14D} in reference to the dense obscuring wind in NGC~5548. High densities have also been shown to be needed as part of `relativistic disc reflection models'  to explain the soft  X-ray excesses seen in   bright Seyfert 1 galaxies, while at the same time allowing for solar iron abundances in the disc atmosphere \citep{2018MNRAS.479..615M, 2019ApJ...871...88G, 2019MNRAS.489.3436J}.

In a `failed wind' scenario \citep[see][for a  review]{2019A&A...630A..94G}, the `classic' BLR emission located at the systemic redshift would then be associated with gas which has failed to reach escape velocity and has fallen back towards the accretion disc at larger radii. Time lags from reverberation mapping campaigns place the `systemic' component of \civ\ emission at $\simeq$50-100 light days, corresponding to $\phi\simeq 10^{20.5}$\,cm$^{-2}$\,s$^{-1}$. We would predict that the outflowing component of \civ\ reverberates with a lag of $\simeq$3\,days, although we note that  the high-luminosity quasars which show the strongest wind signatures are also less variable, and so it is harder to measure robust time lags  in these objects.

In \citet{2020MNRAS.496.2565T}, we identified similarly dense and close-in gas as responsible for emitting the low-ionization \ion{Al}{III}\,$\lambda$1860 and \ion{Fe}{III} ultraviolet lines in quasars, however these species do not show evidence for outflows in their line profiles.
This is consistent with a scenario in which the \ion{Al}{III}\,$\lambda$1860 and \ion{Fe}{III} lines trace the `reservoir' of gas from which the wind is launched.

\section{Conclusions}


Using $\sim$50\,000 absorption-free SDSS quasar spectra with  $z>2$, we have investigated the ratios of the high-ionization rest-frame ultraviolet emission lines \civ, Ly\,$\alpha$, \ion{N}{V}, and the 1400\,\AA\ blend \ion{Si}{IV}$+$\ion{O}{IV}]. 
Our main results are:

\begin{itemize}
    \item The 1216-1255\,\AA\ region  can be well-fit by assuming both Ly\,$\alpha$ and \ion{N}{V} lines have identical kinematic profiles to \civ, suggesting that at least some of the high-ionization ultraviolet line emission in quasar spectra is coming from gas which is outflowing at several thousand \kmps.
    \item The high-ionization line ratios vary strongly and smoothly as a function of their kinematics, suggesting that the  physical conditions of outflowing line-emitting gas are different from the gas which is emitting the `core'  or `systemic' part of the line profile, and that the relative amounts of emission from these two components varies from object to object.
    \item When accounting for the \civ\ outflow signatures, there is no change in the high-ionization line ratios as a function of continuum luminosity, black hole mass, or Eddington fraction.  Previously reported correlations in the literature can be explained by the fact that objects with stronger outflow signatures tend to be more luminous, have larger accretion rates, and have more biased \civ-derived $M_\textrm{BH}$ estimates.
    \item Using photoionization models from \Cloudy, we find that the full observed range of emission-line ratios can be explained by emission from gas at fixed metallicity but with a wide range of density and photon flux.
    \item The observations are consistent with a model in which the high-ionization line  emission is dominated by gas from two distinct regions, where the outflowing (blue) wing of the line is emitted by denser gas, closer in to the central black hole, with $n_\textrm{H}\approx 10^{13-14}\,\textrm{cm}^{-3}$ and $\phi\approx 10^{22-24}\,\textrm{cm}^{-2}\,\textrm{s}^{-1}$. This corresponds to the radius from where line-driven winds are predicted to be launched. 
    The `core' part of the line profile is emitted by gas within the typically assumed range of BLR properties, perhaps in some sort of LOC model:
    $n_\textrm{H}\approx 10^{9-12}\,\textrm{cm}^{-3}$ and $\phi\approx 10^{18-21}\,\textrm{cm}^{-2}\,\textrm{s}^{-1}$, and could correspond to the failed part of the wind.
    \item The metallicity of the line-emitting gas in this model could be solar, or could be within a factor of a few of solar, but need not vary  across the quasar population in order to account for the observed variation in emission line properties. Works using quasar BLR line ratios in galaxy evolution studies therefore must consider changes in the physical conditions of the emitting gas before inferring chemical enrichment histories. 
\end{itemize}

\section*{Acknowledgements}


MJT acknowledges support from Fondo ALMA 31190036, and thanks the IoA, Cambridge, for the award of visitor status while this work was completed.
GJF acknowledges support by NSF (1816537), NASA (ATP 17-ATP17-0141), and STScI (HST-AR-15018).
ALR and PCH acknowledge funding from the Science and Technology Facilities Council.
MC acknowledges support from STScI (HST-AR-14556.001-A), NSF (1910687), and NASA (19-ATP19-0188).
This work  made use of \textsc{astropy} \citep{astropy:2013, astropy:2018}, \textsc{corner.py} \citep{corner}, \textsc{matplotlib} \citep{Hunter:2007}, \textsc{numpy} \citep{numpy}, and \textsc{scipy} \citep{scipy}.
We thank Bob Carswell, Fred Hamann
and James Matthews for useful discussions and encouraging feedback,
and an anonymous reviewer for their thoughtful comments.

Funding for the Sloan Digital Sky Survey IV has been provided by the Alfred P. Sloan Foundation, the U.S. Department of Energy Office of Science, and the Participating Institutions. SDSS-IV acknowledges
support and resources from the Center for High-Performance Computing at
the University of Utah. The SDSS web site is www.sdss.org.

SDSS-IV is managed by the Astrophysical Research Consortium for the 
Participating Institutions of the SDSS Collaboration including the 
Brazilian Participation Group, the Carnegie Institution for Science, 
Carnegie Mellon University, the Chilean Participation Group, the French Participation Group, Harvard-Smithsonian Center for Astrophysics, 
Instituto de Astrof\'isica de Canarias, The Johns Hopkins University, Kavli Institute for the Physics and Mathematics of the Universe (IPMU) / 
University of Tokyo, the Korean Participation Group, Lawrence Berkeley National Laboratory, 
Leibniz Institut f\"ur Astrophysik Potsdam (AIP),  
Max-Planck-Institut f\"ur Astronomie (MPIA Heidelberg), 
Max-Planck-Institut f\"ur Astrophysik (MPA Garching), 
Max-Planck-Institut f\"ur Extraterrestrische Physik (MPE), 
National Astronomical Observatories of China, New Mexico State University, 
New York University, University of Notre Dame, 
Observat\'ario Nacional / MCTI, The Ohio State University, 
Pennsylvania State University, Shanghai Astronomical Observatory, 
United Kingdom Participation Group,
Universidad Nacional Aut\'onoma de M\'exico, University of Arizona, 
University of Colorado Boulder, University of Oxford, University of Portsmouth, 
University of Utah, University of Virginia, University of Washington, University of Wisconsin, 
Vanderbilt University, and Yale University.

\section*{Data Availability}

The data underlying this article were accessed from the Sloan Digital Sky Survey (www.sdss.org).


\bibliographystyle{mnras}
\bibliography{paper_refs.bib}



\appendix

\section{The approach to the black body limit}
\label{sec:bblim}

Perhaps the most surprising result of the photoionization calculations presented in Section~\ref{sec:models} is the fact that the \ion{N}{V} : Ly$\alpha$
line ratio approaches 1:1 at high densities.
This is because emission from the cloud is going over to the 
black body limit in which statistical mechanics rather
than detailed atomic physics or composition determine the intensity of spectral lines.
The spectral synthesis code we use in this work, \Cloudy, is designed to go to the proper
physical limits in well-defined  extremes
(\citealt{2013RMxAA..49..137F}, figs. 17 \& 18;  \citealt{2017RMxAA..53..385F},
figs. 10 \& 11). 
Figures 1, 2, 3, 6 and 7 of  \citet{1988ApJ...332..141F} demonstrate
the approach to LTE,  STE, and the black-body-emission limits.

In this Appendix, we describe the physics behind the results shown in Figures \ref{fig:theory6} 
and \ref{fig:v}.
The principle of micro reversibility in atomic physics or detailed balance in statistical
mechanics is most commonly expressed by the Kirchhoff-Planck Law
\begin{equation}
j_{\nu} = \kappa_{\nu} B_{\nu}\left ( T_\textrm{ex} \right  )
\end{equation}
where $j_{\nu}$ and $\kappa_{\nu}$ are the line emissivity 
[erg cm$^{-3}$ s$^{-1}$ sr$^{-1}$] and opacity [cm$^{-1}$] respectively.
They are related by the Planck function $B_{\nu}$ at the line excitation temperature $T_\textrm{ex}$
by the more typical form of the Kirchhoff-Planck Law 
\begin{equation}
B_{\nu}\left ( T_\textrm{ex} \right  ) = \frac{j_{\nu} }{ \kappa_{\nu} } 
\ [\textrm{erg cm}^{-3} \ \textrm{s}^{-1} \ \textrm{sr}^{-1}] .
\label{eq:KPlaw}
\end{equation}

Considering only line emission from a cloud of thickness $L$ the intensity
emergent from a slab will be
\begin{equation}
    I_{\nu} = \int_0^L j_{\nu} \exp (-\tau_{\nu} ) \ dl =
    \int_0^L  j_{\nu} \exp (-\kappa_{\nu} l ) \ dl
\end{equation}
where $\tau_{\nu}$ is the optical depth. 
This has the solution
\begin{equation}
    I_{\nu} = \frac{ j_{\nu} }{\kappa_{\nu} } \left[ 1-\exp (-\kappa_{\nu} L ) \right].
\end{equation}
In the optically thin limit, $\tau= \kappa_{\nu} L  \ll 1$, the  emission goes
over to the optically thin limit
\begin{equation}
 I_{\nu} = \frac{ j_{\nu} }{\kappa_{\nu} } \left[ 1-\exp (-\kappa_{\nu} L ) \right]
 = \frac{ j_{\nu} }{\kappa_{\nu} } \kappa_{\nu} L = j_{\nu} L
\end{equation}
i.e., the intensity is linearly related to the line emissivity and the cloud thickness.
In the optically thick thermalized limit the intensity becomes
\begin{equation}
    I_{\nu} = \frac{ j_{\nu} }{\kappa_{\nu} } \left[ 1-\exp (-\kappa_{\nu} L ) \right] =
    \frac{ j_{\nu} }{\kappa_{\nu} } = B_{\nu} ,
\end{equation}
i.e., the line has saturated at the blackbody limit and its intensity no longer
depends on atomic physics, abundances, or cloud thickness.
Physically, we ``see'' the Planck function emitted from the surface of the cloud.

The previous discussion shows why the intensity of a thermalized emission line will 
go over to the Planck function in the large optical depth limit.  
The  optical depth is not the whole story.
A slope of forty-five degrees in Figure \ref{fig:theory6} corresponds to 
clouds with the same ionization parameter.
In photoionization equilibrium, the  ion column density $N_\textrm{ion}$ is
related to the flux of ionizing photons and the density $n$ by the balance equation
$\phi($H$)\approx n^2 L \alpha_B$ where $L$ and $\alpha_B$ are the
Stromgren length and recombination rate coefficient respectively 
\citep{2006agna.book.....O}.
The optical depth $\tau$  is proportional to the column density 
so $\tau \propto N \propto \phi/n \alpha_B \propto U$, the ionization parameter.
We see from Figure \ref{fig:theory6} that only the denser clouds in the upper
right quadrant have line ratios approaching unity.
Density matters too, which we discuss next.

This discussion focuses on the formation of strong lines like \ion{N}{V}\,$\lambda$1240 or \ion{C}{IV}\,$\lambda$1549
which can, to a fair approximation, be treated as a two-level system with
upper and lower levels $u,l$.
We neglect background opacities and line photoexcitation, both fair approximations
for these lines in AGN.

The line source function $S_\nu$ describes the interplay between emission $j_\nu$ and absorption
$\kappa_{\nu}$ and can be defined as
\begin{equation}
S_nu = \frac{j_\textrm{scat}+j_\textrm{therm}}{\kappa_{\nu} }
\label{eq:Sdefn}
\end{equation}
where $j_\textrm{scat}$ and $j_\textrm{therm}$ are the emissivities due to scattering and
thermal emission.\footnote{
Chapter 2 of the open-access radiative transfer text \citet{2003rtsa.book.....R}
gives a good introduction to the source function and its relationship with scattering
and thermal emission.}
The absorption term appears in the Kirchhoff-Planck Law (Equation \ref{eq:KPlaw}),
but the scattering part does not since it is a non-thermal process.
It is customary to define the line thermalization probability  as
\begin{equation}
\epsilon_{u,l}= \frac{C_{u,l}}{C_{u,l}+A_{u,l} } =
\frac{q_{u,l} n_\textrm{coll}}{q_{u,l} n_\textrm{coll}+A_{u,l} }
\end{equation}
where $C_{u,l}$ and $A_{u,l}$ are the collision and radiative rates [s$^{-1}$] 
(see \citealt{2019Obs...139..152F}),
$q_{u,l}$ is the collisional rate coefficient [cm$^3$ s$^{-1}$], and
$n_\textrm{coll}$ is the density of colliders [cm$^{-3}$].
The source function (Equation \ref{eq:Sdefn}) can then be written as
\begin{equation}
S_\nu = \frac{j_\textrm{scat}+j_\textrm{therm}}{\kappa_{\nu} }
= \left( 1 - \epsilon_{u,l} \right) J_\nu + \epsilon_{u,l} B_{\nu} 
\label{eq:Sepsilon}
\end{equation}
where the first term on the right expresses the scattering emission in terms of 
the local mean intensity $J_{\nu}$ and the second term is a restatement of
the Kirchhoff-Planck form of thermal emission.

The system goes over to the high-density limit 
when the density is greater than the critical density of colliders $n_\textrm{crit}$,
the density where radiative and collisional de-excitation are
equally probable, $q_{u,l} \ n_\textrm{crit} = A_{u,l}$ \citep{2006agna.book.....O}.
In this limit $\epsilon_{u,l} \rightarrow 1$ and we recover the Kirchhoff-Planck Law,
Equation \ref{eq:Sepsilon} goes to
$S_\nu = B_\nu$.
Large optical depths are not a sufficient condition for the blackbody
limit to be reached, high densities are necessary too.
These high densities are the novel aspect of the analysis done in this paper.

These considerations manifest in several ways in Figures \ref{fig:theory6} and \ref{fig:v}. 
The \ion{N}{V}/Ly$\alpha$ ratio in Figure \ref{fig:theory6} has the greatest
diagnostic value.  Lines with a 45 degree slope correspond to clouds with the
same ionization parameter and line optical depths.  The ratio saturates
at a roughly 1:1 ratio in the upper right quadrant because the lines have
become thermalized and approach the blackbody limit. 
This allows solar-abundance clouds to produce very large \ion{N}{V}/Ly$\alpha$ intensity ratios.
Figure \ref{fig:v} is a series of clouds with the same density and ionization 
parameter but with varying microturbulence.  
Although the species column densities do not change,
the line optical depths decrease linearly as the turbulence increases.
The ratios are all of order unity for smaller turbulence (and larger optical depth) since the spectrum
is closer to the thermal limit.
The ratios fall below unity at larger turbulence as the spectrum desaturates. 

\bsp	
\label{lastpage}
\end{document}